\documentclass[11pt]{amsart}
\usepackage{amsmath}
\usepackage{amsfonts}
\usepackage{amssymb}

\newcommand{\sst}{\scriptscriptstyle}
\newcommand{\dst}{\displaystyle}
\newcommand{\tst}{\textstyle}

\newcommand{\pa}{\partial}

\newcommand{\ra}{\rightarrow}

\newcommand{\ti}{\times}

\newcommand{\fr}[2]{{\textstyle \frac{#1}{#2} }}

\newcommand{\fsl}{{\mathfrak s}{\mathfrak l}}

\newcommand{\hsl}{\mbox{$\widehat{sl_2}$}}

\newcommand{\al}{\alpha}
\newcommand{\be}{\beta}
\newcommand{\ga}{\gamma}
\newcommand{\Ga}{\Gamma}
\newcommand{\de}{\delta}

\newcommand{\ep}{\epsilon}

\newcommand{\om}{\omega}

\newcommand{\si}{\sigma}

\newcommand{\vf}{\varphi}

\newcommand{\bJ}{\bar{J}}

\newcommand{\bn}{\bar{n}}
\newcommand{\bv}{\Bar{v}}

\newcommand{\bx}{\bar{x}}
\newcommand{\bm}{\bar{m}}
\newcommand{\bz}{\bar{z}}

\newcommand{\CC}{{\mathcal C}}
\newcommand{\CD}{{\mathcal D}}

\newcommand{\CF}{{\mathcal F}}

\newcommand{\CH}{{\mathcal H}}

\newcommand{\CO}{{\mathcal O}}  
\newcommand{\CP}{{\mathcal P}}  
  
\newcommand{\CR}{{\mathcal R}}
\newcommand{\CS}{{\mathcal S}}

\newcommand{\BR}{{\mathbb R}}

\newcommand{\BC}{{\mathbb C}}

\newcommand{\BZ}{{\mathbb Z}}

\DeclareMathOperator{\Tr}{Tr}
\newcommand{\LH}{H_3^{+}}
\newcommand{\rf}[1]{(\ref{#1})}
\newcommand{\aufz}
{\begin{list}{$\bullet$}{\topsep0cm \itemsep0cm \parsep0cm}}
\newcommand{\eaufz}{\end{list}}
\newcounter{num}
\newcommand{\remlst}{\begin{list}
{(\arabic{num})}{\usecounter{num}\topsep0cm \itemsep0cm \parsep0cm}}
\begin{document}
\thispagestyle{empty}
\hspace*{\fill} LPM-97/02\\
\hspace*{\fill} hep-th/9712258\\[.5cm]
\title{The mini-superspace limit of the SL(2,C)/SU(2)-WZNW model
\footnote{Work supported by the 
European Union under contract FMRX-CT96-0012}}
\author{J. Teschner}
\address{Laboratoire de Physique Math\'ematique,
Universit\'{e} Montpellier II,
Pl. E. Bataillon, 34095 Montpellier, France\\
teschner@lpm.univ-montp2.fr}

\begin{abstract}
Many qualitatively new features of WZNW models associated to noncompact 
cosets are due to zero modes with continuous spectrum. Insight may be 
gained by reducing the theory to its zero-mode sector, the mini-superspace
limit. This will be discussed in  some detail for the example of
SL(2,C)/SU(2)-WZNW model. The mini-superspace limit of this model can be
formulated as baby-CFT. Spectrum, structure constants and
fusion rules as well as factorization of four
point functions are obtained from the harmonic analysis on SL(2,C)/SU(2). 
The issues of operator-state correspondence or the appearance of 
non-normalizable intermediate states in correlation functions can be
discussed transparently in this context.
\end{abstract}

\maketitle

\section{Introduction}

In recent years there as been a lot of progress in the subject of
rational conformal field theories (RCFT's, finite number of primary fields)
or quasi-RCFT's (infinite number of primary fields but finite-dimensional
fusion or braid relations). Not much is known on a class of theories that
might be called non-compact CFT's: These have continuous families of primary 
fields and the operator product expansion of two primary field will
generically involve an integral over (a subset of) the continuum of 
primary fields available. Such theories are more difficult to study, as i.e.
there are generically no nullvectors
in the relevant current algebra representations, so that most of the 
techniques from rational conformal field theories are not available.  

One of the simplest examples for a noncompact CFT in the above sense is 
Liouville theory. By formal path-integral arguments \cite{S}\cite{P}
one was led there to expect more qualitatively new features
as compared to (quasi)-RCFT's:
\begin{enumerate}
\item Loss of one-to-one correspondence between states and operators, which 
is a fundamental axiom in many approaches to RCFT. One aspect of this 
issue is that
one will have to distinguish between operators corresponding to
normalizable and non-normalizable states respectively. 
\item Conditional nature of factorization: The set of intermediate 
represenations to factorize over will in general not coincide with the 
spectrum of the theory. Although outside the spectrum, 
it may happen that non-normalizable states appear in intermediate channels 
of correlation function. 
\end{enumerate}

Most of these points are related to the fact that one has zero modes with 
continuous spectrum. It is therefore useful to consider a limiting case
in which only the zero mode dependence survives. Such a limiting case for
Liouville theory has been discussed under the name of mini-superspace limit 
in \cite{S}. It may be understood as considering only space-independent field
configurations when spacetime is a cylinder.

Another simple example that has been studied i.e. in view of applications
to the stringy eucliedean 2D black hole is the WZNW-model 
corresponding to the coset $H_3^+\equiv SL(2,\BC)/SU(2)$, cf. \cite{Ga}
and references therein. The mini-superspace limit of the 2D euclidean black
hole CFT which can be easily obtained from the mini-superspace limit 
of $H_3^+$ WZNW model has been used to obtain amplitudes for reflection
of strings in the 2D euclidean black hole geometry in \cite{DVV}.

My aim will be to develop the mini-superspace limit of the $H_3^+$-WZNW model 
in some detail. This turns out to be just quantum mechanics with configuration
space $H_3^+$, mathematically nothing but harmonic analysis on that
symmetric space. Some aspects of this limit have been discussed in 
\cite{Ga}, but the present discussion will be quite complimentary to that
given in loc. cit.. 

One purpose is to illustrate how the above mentioned new qualitative
features of noncompact CFT's are naturally understood in analogy with
the harmonic analysis on symmetric spaces. To this aim a formulation will be 
presented that makes an analogy with the bootstrap approach of \cite{BPZ}
transparent.

However, the mini-superspace limit serves not only for illustration
of qualitative features: It is expected to
describe a certain semiclassical limit of the full theory, so one may use it
to check the exact results for the full theory proposed in \cite{T1}.
This works both for the structure constants and the fusion rules.

Furthermore, it was proposed in \cite{MSS} that in the context of
2D quantum gravity or noncritical string theory the mini-superspace limit
may even be exact. Similar assumptions were used in \cite{DVV} to
obtain the reflection amplitudes of strings in the 2D euclidean black 
hole geometry. Let me however note that such reflection amplitudes can now be
compared to exact results proposed in
\cite{T1}, where one finds explicit quantum corrections to the 
mini-superspace result. 

The present paper is the first in a series of papers devoted to the study 
of the $H_3^+$ and $SL(2,\BR)$-WZNW models. The next paper \cite{T1}
contains a derivation of an exact expression for the structure constants
by using the methods of \cite{TL},
from which one can find a reflection amplitude as in \cite{ZZ}. It further
suggests a way to obtain spectrum and fusion rules from 
canonical quantization.

Some mathematical foundations on the use of non-highest or lowest weight
representations to construct conformal blocks 
are laid in \cite{T2}, which treats the $SL(2,\BR)$ case on equal footing.
This will be used to give a treatment of the $SL(2,\BR)$ case along the
lines of \cite{T1} in a forthcoming publication.   

The contents of the present paper are as follows: The second section
discusses quantization of the mini-superspace limit: Space of
states and operators, where the momentum operators in a Schr\"{o}dinger
representation can be chosen to represent the Lie-algebra of the 
symmetry group $SL(2,\BC)$, whereas primary fields can be realized as 
multiplication operators. The issue of operator-state correspondence 
can be made quite transparent in this context.

The third section studies correlation functions. First it is described
how far one can get by exploiting $SL(2,\BC)$ symmetry in a bootstrap 
appraoch \`{a} la \cite{BPZ}. This treatment of the mini-superspace limit
as baby-CFT makes the structural analogy to the full theory transparent.
This is afterwards compared to the definition and (in some cases)
calculation of correlation functions as overlaps. Fusion rules are 
here obtained by relating them to the spectral decomposition. It is
explained how contributions of non-normalizable intermediate states arise
in a well defined manner.

Two appendices contain certain technical aspects: Appendix A treats the 
spectral decomposition of the Laplacian of $H_3^+$. This could have 
been extracted from \cite{GGV}, but since it would also have taken some
time to explain how all the results needed here follow from those given
there, I preferred to give a different treatment, self-contained and
adapted to the present needs. 

Appendix B shows how one may explicitely calculate 
fusion relations for the mini-superspace conformal blocks.

\section{The mini-superspace limit of $H_3^+$-WZNW-model}

In a Lagrangian formulation, one may formulate the $H_3^+$-WZNW-model by
starting from an $SL(2,\BC)$ model and gauging the $SU(2)$ subgroup, 
see i.e. \cite{Ga} and references therein. Equivalently one may realize the
coset $SL(2,\BC)/SU(2)$ as the space of two-by-two hermitian complex matrices 
$h$ with unit determinant and start from the action \cite{Ga}
\begin{equation}
S[h]=\frac{1}{\pi}\int d^2z\Bigl((\pa_z\psi)(\pa_{\bz}\psi)+
(\pa_z+\pa_z\psi)\bv(\pa_{\bz}+\pa_{\bz}\psi)v\Bigr)
\end{equation} 
where $h$ was parametrized as
\begin{equation} h=\left(\begin{array}{cc} 
e^{\psi}(1+|v|^2) & v \\
\bv & e^{-\psi}\end{array}\right). \end{equation}
If one considers the theory on a cylinder with periodic space and infinite
time, and furthermore
restricts to field configurations that are independent of the space variable, 
one gets the action
\begin{equation}
S_m[h]=\frac{\kappa}{4}\int dt \Tr\bigl(h^{-1}\pa_t h\bigr)^2
\end{equation}
This action is real and invariant under the $SL(2,\BC)$-symmetry
$h\ra g^{-1}h(g^{-1})^{\dagger}$. One therefore expects this
symmetry to be unitarily
realized in the corresponding quantum theory. 

\subsection{Space of states}

The Schr\"{o}dinger representation for the quantum mechanics with phase space 
$T^*H_3^+$ is obtained by taking the Hilbert space to consist of functions
on $ H_3^+$, square-integrable w.r.t. the measure $dh=d\phi d^2v$
if $h$ is parametrized as 
\begin{equation}\label{param} h=\left(\begin{array}{cc} 
e^{\phi}\sqrt{1+|v|^2} & v \\
\bv & e^{-\phi}\sqrt{1+|v|^2} \end{array}\right). \end{equation}
The symmetry group $SL(2,\BC)$ acts on wave-functions on $ H_3^+$ via
\begin{equation} 
\label{trdef}T_g \Psi(h)=\Psi(g^{-1}h (g^{-1})^{\dagger}).\end{equation} 
The point about the present choice of scalar product is that it realizes 
the $SL(2,\BC)$-symmetry {\it unitarily}:

This may be seen by noting that each $h\in H_3^+$ can be written as
$h=gg^{\dagger}$ for some $g\in SL(2,\BC)$. The point is to observe that
\begin{equation}\label{intid} 
<\Psi_2,\Psi_1> \equiv \int_{H_3^+}dh \Psi^*_2(h)\Psi_1(h)=
V_{SU(2)}^{-1}\int_{SL(2,\BC)}dg \Psi^*_2(gg^{\dagger})\Psi_1(gg^{\dagger})
, \end{equation}
where dg is the $SL(2,\BC)$-invariant measure
\[ dg=\left(\frac{i}{2}\right)^4 d^2\al d^2 \be d^2 \ga d^2 \de\;\, 
\de^2(\al\de-\be\ga)
=\left(\frac{i}{2}\right)^3|\al|^{-2} d^2\al d^2\be d^2 \ga\qquad \mbox{if}
\quad g=\left(\begin{array}{cc} \al & \be \\
                        \ga & \de  \end{array}\right),
\]
and $ V_{SU(2)}$ is the volume of $SU(2)$. Given (\ref{intid}), unitarity
of the $SL(2,\BC)$-action is a trivial consequence of the invariance 
property $d(g_0g)=dg$ of the measure $dg$ on $SL(2,\BC)$. In order to
establish (\ref{intid}) one may rewrite the right hand side with the help
of the Iwazawa decomposition 
\[ g=kan=k\left(\begin{array}{cc} e^p & 0 \\ 0 & e^{-p} \end{array} \right)
\left(\begin{array}{cc} 1 & z \\ 0 & 1 \end{array} \right)
\qquad k\in\mbox{SU(2)},\quad z\in\BC,\quad p\in\BR. \]
The measure $dg$ factorizes in this parametrization as $dg=dkdpd^2z$, as 
may be seen by explicit calculation of the Jacobian of the 
change of variables from $(a,b,c)$ to $(k,p,z)$. The integration 
over $SU(2)$ factors out since the integrand is $k$-independent. The
remaining variables $(p,z)$ provide an alternative parametrization
of $H_3^+$. Changing variables from $(p,z)$ to $(\phi,v)$ one finds 
$dpd^2z=dh=d\phi d^2v$ as required. 

It will be important to know the Hilbert space decomposes into $SL(2,\BC)$
irreducible representations. 
This was first found in \cite{GGV}. In order to have a self-contained account
of all the results needed here, I have summarized an alternative approach based
on spectral analysis of the Laplacian on $H_3^+$ in the appendix A.
There one can find (scetches of) the proofs of all the statements made in
this subsection.
 
Abstractly the decomposition reads
\begin{equation}\label{abstdecom} \CH\equiv L^2(H_3^+,dh)=
\int_{\rho>0}^{\oplus} d\rho \rho^2\;\,\CH_{-\frac{1}{2}+i\rho} ,
\end{equation}
where $ \CH_j$ is a representation of the princial series of $SL(2,\BC)$,
which may i.e. be explicitely realized on $L^2(\BC)$ via
\begin{equation}\label{prser}
 T_g f(z)=|\be z + \de|^{4j} f\left(\frac{\al x+\ga}{\be x+\de}\right)
\quad\mbox{if}\quad
g=\left(\begin{array}{cc} \al & \be \\ \ga & \de \end{array} \right). 
\end{equation}

The decomposition (\ref{abstdecom}) may be realized explicitely as 
a kind of Fourier-transform: The Fourier-components are defined by
\[ F(j;x,\bx)\equiv \int_{H_3^+}dh \;\Psi(j;x,\bx|h) \; f(h), \] 
where the kernel $\Psi^j(x,\bx|h)$ that takes the role of the plane waves
$e^{ikx}$ in the ususal Fourier-transform is given as
\[ \Psi(j;x,\bx|h)=\frac{2j+1}{\pi}
\left( (x,1)\cdot h\cdot \binom{\bx}{1}\right)^{2j} .
\] 
It is easy to check from the definitions that $F(j;x,\bx)$ indeed transforms
under $SL(2,\BC)$ as in (\ref{prser}) if $f(h)$ is transformed by
(\ref{trdef}).
 
The function $f(h)$ is recovered from its transform $F(j;x,\bx)$ via the
inversion formula
\[ f(h)=\frac{i}{(4\pi)^3}\int_{\CP_+} dj \;(2j+1)^2
\int d^2x \; \Psi^*(j;x,\bx|h)\; F(j;x,\bx) \qquad\text{where }
\CP_+=-\fr{1}{2}+i\BR^{>0} \]
One may therefore consider the set of functions $\{\Psi(j;x,\bx|.); 
x\in\BC,j\in\CP_+\}$ as a plane wave basis for $\CH$.
Indeed it may be shown (Appendix A)
that they are $\de$-function normalizable:
\begin{equation} \label{norm}
<\Psi(j;x,\bx),\Phi(j';x',\bx')>=2\pi\de^{(2)}(x-x')\de(j-j') 
\qquad\text{for }j,j'\in\CP_+
\end{equation}

The functions $\Psi(j;x,\bx|h)$ and $\Psi(-j-1;x,\bx|h)$ are linearly 
related to each other:
\begin{equation} \Psi(j;x,\bx)=\frac{2j+1}{\pi}\int\! d^2x'\; |x-x'|^{4j}
\Psi(-j-1;x',\bx')\label{reflec} \end{equation} 
The general form of this relation (but not the $j$-dependent prefactor)
is determined by $SL(2,\BC)$-symmetry: The integral operator with kernel 
$|x-x'|^{4j}$ is just the intertwining operator \cite{GGV} 
between representations with spin $-j-1$ and $j$ expressing the 
equivalence of these representations. 

It is sometimes useful to also use an alternative basis $\Psi^j_{np}(h)$
$n\in\BZ$, $p\in\BR$ which is related to the  $\Psi(j;x,\bx|h)$ by the 
following Fourier-transform on $L^2(\BC)$:
\begin{equation} \Psi^j_{np}(h)\equiv
\int_{\BC}d^2x \;\; e^{in\arg(x)}|x|^{-2j-2+ip} \Psi(j;x,\bx|h). \end{equation}
The explicit expression for $\Psi^j_{np}(h)$ in terms of the hypergeometric
function can be found in appendix A.
 
\subsection{Momentum operators: The Lie algebra of $SL(2,\BC)$}

The action of the Lie algebra of $SL(2,\BC)$
on differentiable function on $H_3^+$ is 
given by the differential operators
\begin{equation} \label{Liedef1}\begin{array}{c}\dst
K^af(h)\equiv \left(\frac{d}{dt}f
\left(e^{-tT_a}h e^{-tT_a^{\dagger}}\right)\right)_{t=0}
\qquad 
L^af(h)\equiv \left(\frac{d}{dt}
f\left(e^{-itT_a}h e^{itT_a^{\dagger}}\right)\right)_{t=0}\\[2ex]
T_+=\left(\begin{array}{cc} 0 & 0 \\ 1 & 0 \end{array}\right)\qquad\qquad
T_0=\frac{1}{2}
    \left(\begin{array}{cc} 1 & 0 \\ 0 & -1 \end{array}\right)\qquad\qquad
T_-=\left(\begin{array}{cc} 0 & 1 \\ 0 & 0\end{array}\right) \end{array}
\end{equation}
Alternatively one may use the holomorphic (resp. antiholomorphic) differential
operators
\begin{equation} \label{Liedef2}
J^af(h)\equiv \left(\frac{\pa}{\pa\tau}f
\left(e^{-\tau T_a}h e^{-\bar{\tau} T_a^{\dagger}}\right)\right)_{\tau=0}
\qquad 
\bJ^af(h)\equiv 
\left(\frac{\pa}{\pa\bar{\tau}}f
\left(e^{-\tau T_a}h e^{-\bar{\tau} T_a^{\dagger}}\right)\right)_{\tau=0}
\end{equation}
In terms of the parametrization (\ref{param}) one finds
\begin{equation}
\begin{array}{c}\dst J^+=-e^{-\phi}\sqrt{1+|v|^2}\frac{\pa}{\pa v}-\frac{1}{2}
e^{-\phi}\frac{\bv}{\sqrt{1+|v|^2}}\frac{\pa}{\pa \phi}  \\[1ex]
\dst J^-=-e^{\phi}\sqrt{1+|v|^2}\frac{\pa}{\pa \bv}+\frac{1}{2}
e^{+\phi}\frac{v}{\sqrt{1+|v|^2}}\frac{\pa}{\pa \phi}
\end{array}\qquad\quad J^0=\frac{1}{2}\left(
-v\frac{\pa}{\pa v}+\bv\frac{\pa}{\pa \bv}-
\frac{\pa}{\pa \phi}\right), \end{equation}
whereas the $\bJ^a$ are given by the complex conjugate operators. 
Their hermiticity properties with respect to the $L^2(H_3^+,dh)$ 
scalar product are $(J^a)^{\dagger}=-\bJ^a$, $a=-,0,+$.

The action of these generators
on the Fourier transform $F[f](j,x,\bx)$ of $f(h)$ is then given by 
\begin{equation}\begin{array}{c}
\pi_j(J^a)F[f](j,x,\bx)\equiv F[J^af](j,x,\bx)=\CD^a_j F[f](j,x,\bx)\\[1ex]
\CD^+_j=-x^2\pa_x-2jx \qquad \CD^0_j=x\pa_x-j \qquad \CD^-_j=\pa_x
\end{array}\end{equation}

\subsection{Primary fields, Operator-State correspondence}

In conformal field theory one is interested mostly in the so-called 
primary fields, operators that have a particularly simple transformation
law under the chiral algebra, in the 
case of the $H_3^+$-WZNW model two copies of the Kac-Moody algebra
$\hsl$ generated by the modes $J_n^a$, $\bar{J}_n^a$.
Quite generally the primary fields $\Phi(v|z,\bz)$ may be labelled 
by vectors $v$ in some representation $V$ of the {\it zero-mode subalgebra}
of the chiral algebra which is generated by the  $J^a\equiv J_0^a$, 
$\bar{J}^a\equiv \bar{J}_0^a$:
\begin{equation}\label{cov} 
[J^a_n, \Phi(v|z,\bz)]=z^n  \Phi(\pi_{\sst V}(J^a_0)v|z,\bz), \end{equation}
and an analogous formula for the $\bar{J}_n^a$, 
where $\pi_V(J^a_0)$ denotes the operator that represents $J^a$ on $V$.
If one i.e. takes $V$ to be the irreducible representation realized in 
$\CH_j$ by means of the differential operators $\CD^a_j$ this reads
\begin{equation}\label{covexpl} 
[J^a_n, \Phi^j(x,\bx|z,\bz)]=
z^n  \CD_j^a\Phi^j(x,\bx|z,\bz), \end{equation}
This kind of transformation law is a sufficient condition for 
$\Phi(v|z,\bz)$ to correspond to a primary state $v$, 
i.e. to a state that satisfies $J_n^a v = 0$ for $n>0$ via the usual
operator-state correspondence
\begin{equation}\label{opstat} \lim_{z\ra 0} \Phi(v|z,\bz)|0\!>. 
\end{equation} 
However, a crucial difference between WZNW models corresponding to 
compact resp. non-compact groups is that the zero-mode representations
of the latter are generically infinite-dimensional. This means that the
vector obtained via (\ref{opstat}) is by no means guaranteed to be
normalizable (not even in $\de$-function sense!)\footnote{This fact was 
first pointed out in the context of Liouville theory by Seiberg \cite{S}
and Polchinski \cite{P}.}.

In the presently discussed mini-WZNW model one may see rather explicitly
that indeed one has to distinguish between normalizable and non-normalizable
states. This fact will also be of crucial importance for understanding
the fusion rules.

First note that condition (\ref{covexpl}) 
in the mini-WZNW case reduces just to 
the statement of covariance under the zero-mode subalgbra, the $z$-dependence
disappears. It is easy to 
find such operators in the presently used Schr\"{o}dinger-representation:
These are just the multiplication operators $\Phi^j(x,\bx)$:
\begin{equation} \label{op-stat}  
\Phi^j(x,\bx)\Psi(h):=\Psi(j;x,\bx|h)\Psi(h). 
\end{equation}

In order to speak of operator-state correspondence one needs to define 
the ``vacuum'' $|0\!>$. Its defining property is usually taken to be 
the invariance under the chiral algebra, here the Lie algebra of $SL(2,\BC)$.
The trivial representation of $SL(2,\BC)$ corresponds to $j=0$. The wave 
function of $|0\!>$ is $\Phi(j=0;x,\bx|h)=1$, 
so the operator corresponding to it via (\ref{op-stat}) is just the 
unity operator. However, since the norm of $|0\!>$ is thereby the
(infinite) volume of $H_3^+$, the state  $|0\!>$ is clearly not contained
in the spectrum. 
But still one has the fact that any multiplication operator that 
acts by multiplication with a normalizable function on $L^2(H_3^+,dh)$ 
does create reasonable states from the ``vacuum'' $|0\!>$.
The state $<\!0|$ conjugate to $|0\!>$ is of course a functional well
defined on a suitable subspace of $L^2(H_3^+,dh)$.

\section{Correlation functions}

Since my intention is to present the $H_3^+$ quantum mechanics as a baby 
conformal field theory, I will start by discussiong how far one can get 
by a strategy analogous to the conformal bootstrap of \cite{BPZ}. As in 
the case of full-fledged CFT one will find that the symmetries determine 
the correlation functions to a large extend, but structure constants and
fusion rules
are not easy to determine explicitely within this approach.

The following subsection then explains how in the presently considered baby 
CFT all the missing information can be found by exploiting the harmonic 
analysis on $H_3^+$.

\subsection{Baby-bootstrap}

According to the previous discussion one may try to define vacuum expectation 
values 
\[ <\!0|\Phi^{j_n}(x_n,\bx_n)\ldots\Phi^{j_1}(x_1,\bx_1)|0\!>,\]
where the ``vacuum'' $|0\!>$  is to denote the $SL(2,\BC)$ invariant ``state''.
The $SL(2,\BC)$ invariance of $|0\!>$ then results in a set of differential 
equations 
\[ \sum_{i=1}^{\infty} \CD_{x_i,j}^a   
<\!0|\Phi^{j_n}(x_n,\bx_n)\ldots\Phi^{j_1}(x_1,\bx_1)|0\!>=0,\qquad a=-,0,+, \]
which allow to express the correlation function in terms of its values for 
(say) $x_1=0$, $x_2=1$, $x_n=\infty$. In particular, this determines the two- 
and three point functions almost completely :
\begin{equation}\label{twothree} \begin{array}{rcl}
\multicolumn{2}{l}{<\!0|\Phi^{j_2}(x_2,\bx_2)\Phi^{j_1}(x_1,\bx_1)|0\!>=} 
& \\[1ex]
& \multicolumn{2}{l}{ \qquad\qquad =N(j_1)\;\de(j_2+j_1+1)\de^{(2)}(x_2-x_1)
+B(j)\;\de(j_2-j_1)|x_2-x_1|^{4j_1} } \\[2ex]
\multicolumn{2}{l}{<\!0|\Phi^{j_3}(x_3,\bx_3)\Phi^{j_2}(x_2,\bx_2)
\Phi^{j_1}(x_1,\bx_1)|0\!>=} & \\[1ex]
& \multicolumn{2}{l}{ \qquad\qquad =C(j_3,j_2,j_1)
|x_3-x_2|^{2(j_3+j_2-j_1)}|x_3-x_1|^{2(j_3+j_1-j_2)}|x_2-x_1|^{2(j_2+j_1-j_3)}
} \end{array} \end{equation}
The only thing that may look somewhat unusual to anyone familiar with CFT \`a
la BPZ is the term with $\de^{(2)}(x_2-x_1)$ in the expression for the 
two point function. In real CFT this would be called a contact term. Here it 
is just the term that gives the scalar product \rf{norm} when 
$j_1,j_2$ are restricted to $\CP_+$. 
 
Now the primary fields form multiplets under the symmetry algebra
$\fsl(2,\BC)_L\oplus \fsl(2,\BC)_R$ generated by the holomorphic
(resp. antiholomorphic generators $J^a$ (resp. $\bJ^a$). In order to keep 
the analogy with \cite{BPZ} as close as possible it is natural to introduce
as secondary fields the derivatives 
\[ \Phi^{j,n,\bn}(x,\bx)\equiv \frac{\pa^n}{\pa x^n}
\frac{\pa^{\bn}}{\pa \bx^{\bn}}\Phi^{j}(x,\bx). \]

The general strategy of the bootstrap amounts to construction
of $(n>3)$-point functions in terms of two- and three point functions. This 
will be possible if there are operator product expansions of two 
operators for their arguments close to each other:
\begin{eqnarray}\label{OPE1}
\lefteqn{\Phi^{j_2}(x_2,\bx_2)\Phi^{j_1}(x_1,\bx_1)}\\
&=& \int dj \;\;|x_2-x_1|^{2(j_2+j_1-j)}\sum_{n,\bn=0}^{\infty}
\CC_{n\bn}(j;j_2,j_1)(x_2-x_1)^n(\bx_2-\bx_1)^{\bn}
\;\Phi^{j,n,\bn}(x_1,\bx_1) 
\nonumber\end{eqnarray}
As in the case of real CFT the requirement that both sides of \rf{OPE1}
transform the same way under the symmetry algebra allows to fix the 
coefficients $ \CC_{n\bn}(j;j_2,j_1)$ uniquely in terms of $ C(j;j_2,j_1)
\equiv \CC_{00}(j;j_2,j_1)$. In the present baby CFT it is easily possible
to carry this out explicitly:
\begin{equation}\label{CCexp}
\begin{array}{rcl} \CC_{n\bn}(j;j_2,j_1)&=& R_n(j;j_2,j_1)R_{\bn}(j;j_2,j_1)
D(j;j_2,j_1) \\[1ex]
R_n(j;j_2,j_1) & = &\dst
\frac{\Ga(j_1-j_2-j-1+n)\Ga(-2j-1-n)}{\Ga(j_1-j_2-j-1)\Ga(-2j-1)n!}
 \end{array}\end{equation}
If one knew both the range of values for $j$ 
in \rf{OPE1}, i.e. the fusion rules, and the explicit expression for
the structure constants $ D(j;j_2,j_1)$ then one could in principle 
unambigously determine any $n>3$-point function: Inserting \rf{OPE1} i.e.
into a four point function leads to the expansion 
\begin{equation}\label{s-channel}\begin{split}
<\Phi^{j_4}\ldots \Phi^{j_1}>=& \int dj_{21}\; 
|x_2-x_1|^{2(j_2+j_1-j_{21})}
\sum_{n,\bn=0}^{\infty}
(x_2-x_1)^n(\bx_2-\bx_1)^{\bn} \\
 & \CC_{n\bn}(j_{21};j_2,j_1)<\!0|\Phi^{j_4}(x_4,\bx_4)
\Phi^{j_3}(x_3,\bx_3)\Phi^{j_{21},n,\bn}(x_1,\bx_1)|0\!>
\end{split}\end{equation}
By observing that the x-dependent pieces factorize into parts depending
holomorphically resp. anti-holomorphically on the $x_i$ one may  
cast the expansion into the form
 \begin{equation}\label{s-chexp}
\begin{array}{rcl}
<\Phi^{j_4}\ldots \Phi^{j_1}>&=& \dst \int dj_{21} C(j_4,j_3,j_{21})
D(j_{21};j_2,j_1)
\Big|\CF_{j_{21}}^s \Big[\,
{}^{j_4}_{x_4}{}^{j_3}_{x_3}{}^{j_2}_{x_2}{}^{j_1}_{x_1}\Big]\Big|^2
\end{array}\end{equation} 
where the (s-channel) ``conformal blocks'' $\CF_{j_{21}}^s$ are defined as 
\begin{equation}\begin{split}
\CF_{j_{21}}^s \Big[\,
{}^{j_4}_{x_4}{}^{j_3}_{x_3}{}^{j_2}_{x_2}{}^{j_1}_{x_1}\Big]
= & \sum_{n=0}^{\infty}R_n(j_{21}|j_2,j_1)
\;\,\frac{\pa^n}{\pa x^n}C\left(\begin{smallmatrix}
j_4 & j_3 & j_{21} \\
x_4 & x_3 & x \end{smallmatrix}\right)
C\left(\begin{smallmatrix}
j_{21} & j_2 & j_1 \\
x & x_2 & x_1 \end{smallmatrix}\right)\\
\text{where}\quad C\left(\begin{smallmatrix}
j_3 & j_2 & j_1 \\
x_3 & x_2 & x_1 \end{smallmatrix}\right)=&
(x_3-x_2)^{j_3+j_2-j_1}(x_3-x_1)^{j_3+j_1-j_2}(x_2-x_1)^{j_2+j_1-j_3}
\end{split}\end{equation}
The sum may be carried out explicitely in terms of the hypergeometric
function: Let $x$ be
the crossratio $x=\frac{(x_2-x_1)(x_4-x_3)}{(x_3-x_1)(x_4-x_2)}$, 
\begin{equation}
\begin{split}
\CF_{j_{21}}^s \Big[\,
{}^{j_4}_{x_4}{}^{j_3}_{x_3}{}^{j_2}_{x_2}{}^{j_1}_{x_1}\Big]=&
(x_4-x_3)^{j_4+j_3-j_2-j_1}(x_4-x_2)^{2j_2}(x_4-x_1)^{j_4+j_1-j_3-j_2} \\
& (x_3-x_1)^{j_3+j_2+j_1-j_4} x^{j_1+j_2-j_{21}}
F(j_4-j_3-j_{21},j_1-j_2-j_{21};-2j_{21};x).\end{split}
\end{equation}
The decomposition \rf{s-chexp}
makes explicit to which extend the correlation functions 
are determined by the symmetry: The conformal blocks are completely determined
by it, whereas one has without further input no information on 
structure constants $C(j_3,j_2,j_1)$ and fusion rules
(range of integration over $j_{21}$).

The additional requirement that one may expect to determine these 
pieces of information also is  
crossing symmetry: One may use an expansion of type
\rf{OPE1} for the product of operators $\Phi^{j_3}\Phi^{j_2}$ instead 
to get an expansion of the four point function into a different 
set of conformal 
blocks (t-channel):
\begin{equation}\label{t-chexp}
\begin{array}{rcl}
<\Phi^{j_4}\ldots \Phi^{j_1}>&=& \dst \int dj_{32} C(j_4,j_{32},j_1)
D(j_{32};j_3,j_2)
\Big|\CF_{j_{32}}^t  \Big[
{}^{j_4}_{x_4}{}^{j_2}_{x_2}{}^{j_3}_{x_3}{}^{j_1}_{x_1}\Big]\!\Big|^2 ,
\end{array}\end{equation}
The result should of course be equal to
the expansion \rf{s-chexp}. Equality of the two expansions (crossing symmetry)
leads to restrictions for the structure constants and fusion rules:

This may be made more explicit by observing that one has  
fusion transformations for the conformal blocks in this 
baby CFT, as shown in Appendix B. They take the form
\begin{equation}\label{deffus}
\CF_{j_{21}}^s \Big[\,
{}^{j_4}_{x_4}{}^{j_3}_{x_3}{}^{j_2}_{x_2}{}^{j_1}_{x_1}\Big]\!=\dst
\int d\mu(j_{32})\;\; F_{j_{21}j_{32}}\!\!\left[{}^{j_3j_2}_{j_4j_1}\right]
\CF_{j_{32}}^t  \Big[\,
{}^{j_4}_{x_4}{}^{j_2}_{x_2}{}^{j_3}_{x_3}{}^{j_1}_{x_1}\Big]\!
\end{equation}
Given fusion relations \rf{deffus}, the requirement of crossing
symmetry translates 
itself into a system of equations for the structure constants:
\begin{equation}\label{cross-gen}
\begin{split}
\dst\int_{\CS_s} d\mu(j_{21}) 
\;F_{j_{21}^{}j_{32}^{}}\!\!\left[{}^{j_3j_2}_{j_4j_1}\right]
\bar{F}_{j_{21}^{}j_{32}'}\!\!\left[{}^{j_3j_2}_{j_4j_1}\right]
\;  & C(j_4,j_3,j_{21})D(j_{21};j_2,j_1) \\
= \delta(j_{32}^{},j_{32}')\; & C(j_4,j_{32},j_1)
C(j_{32};j_3,j_2).
\end{split}\end{equation}
Instead of solving this horrible system of equations one has in the 
present baby CFT a direct way to obtain structure constants and 
fusion rules, which will be discussed next.

\subsection{N-point functions as overlaps}

According to the previous discussion of state-operator correspondence one may
alternatively define vacuum expectation values 
with help of the scalar product in $L^2(\LH,dh)$:
\begin{equation} <\Phi^{j_{\sst N}}(x_{\sst N},\bx_{\sst N})
\ldots\Phi^{j_1}(x_1,\bx_1)>=
\int_{\LH}dh\prod_{i=1}^N \Psi(j_i;x_i,\bx_i|h) 
\end{equation}
In order to discuss convergence of such integrals it is easier to 
consider
\[ <\Phi^{j_{\sst N}}_{n_{\sst N},p_{\sst N}}
\ldots\Phi^{j_1}_{n_1p_1}>=
\delta({\tst \sum n_i})\delta({\tst \sum \om_i})\int_0^{\infty}dy
\prod_{i=1}^N\Theta^{j_i}_{n_ip_i}(y), \]
where
\begin{eqnarray*}
\Theta^{j}_{np}(y)&=&B^{-1}_{np}(j)y^{|n|}(1+y)^{ip}\;
F\Big(\fr{1}{2}(|n|+ip)-j,\fr{1}{2}(|n|+ip)+j+1;1+|n|;-y\Big) \\
& & B_{np}(j)\equiv \frac{\Ga(1+|n|)\Ga(2j+1)}{\Ga(\frac{1}{2}(|n|+ip)+j+1)
\Ga( \frac{1}{2}(|n|-ip)+j+1)}. 
\end{eqnarray*} 
In order to find the conditions for convergence of the integral
over $y$ one needs the asymptotics of $\Theta^{j}_{m\bm}(y)$ for 
$y\ra\infty$:
\[ \Theta^{j}_{np}(y)\sim y^{-j-1}+ 
y^j \frac{B_{np}(j)}{B_{np}(-j-1)}, \]
The overlap defining the n-point functions will therefore be convergent 
provided
\[ \sum_{i=1}^{n} \tst
\left( \left|\Re\left(j_i+\frac{1}{2}\right)\right|-\frac{1}{2}
\right)<-1 \]
One observes that this condition can never be satisfied for $n\neq 2$, 
a case which
requires special discussion. However, it is important to observe that these
integrals give well defined correlation functions with $n>2$ not only 
for operators corresponding to normalizable states $\Re(j_i)=-1/2$ but also
for a class of operators corresponding to non-normalizable states.

To finish this subsection, I would like to make the following remark:
Since the general classical solution of Liouville theory in the case of only 
``weak'' insertions is given by 
\[ e^{-j\phi_L(z,\bz)}=
\left( (z,1)\cdot gg^{\dagger} \cdot \binom{\bz}{1}\right)^{2j_i}
\qquad g\in SL(2,\BC) \]
one finds correspondence between WZNW and Liouville semiclassical 
correlation function via $x\ra z$.

\subsubsection{Two-point function} 

The two-point function of operators $\Phi^{j_2}$, $\Phi^{j_1}$ with
$j_i\in -\fr{1}{2}+i\BR$, $i=1,2$ 
may be read off from the orthogonality relations
found in the Appendix as 
\[ <\Phi^{j_2}_{n_2p_2}\Phi^{j_1}_{n_1p_1}>=
(2\pi)^3 \delta_{n_1+n_2}\delta(p_2+p_1)\left(
i\delta(j_1+j_2+1)+\frac{B_{n_1p_1}(j_1)}{B_{n_1p_1}(-j_1-1)}
i\delta(j-j') \right). \]
In terms of the $\Phi^j(x,\bx)$ it reads
\[ <\Phi^{j_2}(x_2,\bx_2)\Phi^{j_1}(x_1,\bx_1)>=2\pi
\left(\de^{(2)}(x_2-x_1)i\de(j_2+j_1+1)
+|x_2-x_1|^{4j_1}\frac{2j_1+1}{\pi}i\de(j_1-j_2)\right). \]
There does not seem to be a way to extend this result to $j_i$ with
$\Re(2j_i+1)\neq 0$. 

\subsubsection{Three point function} 

The integral defining the three point function of operators 
$\Phi_i=\Phi^{j_i}(x_i,\bx_i)$ $i=1,2,3$ has been
calculated in \cite{ZZ} with the result 
\begin{eqnarray*}
<\Phi_3\Phi_2\Phi_1>&=&\pi^{-3}
\frac{\Ga(-j_1-j_2-j_3-1)\Ga(j_3-j_2-j_1)\Ga(j_2-j_1-j_3)
\Ga(j_1-j_2-j_3)}{\Ga(-2j_1-1)\Ga(-2j_2-1)\Ga(-2j_3-1)}\\
& &\times
|x_3-x_2|^{2(j_3+j_2-j_1)}|x_3-x_1|^{2(j_3+j_1-j_2)}|x_2-x_1|^{2(j_2+j_1-j_3)}
\end{eqnarray*}
Note that in contrast to the two point function it is obviously possible to 
meromorphically continue the result to general values of the $j_i$. 

\subsection{Operator product expansions}

According to the previous discussion on  operator-state correspondence 
the operator obtained by taking the product of two operators 
$\Phi^{j_2}(x_2,\bx_2)$ and 
$\Phi^{j_1}(x_1,\bx_1)$ is just represented by the product of the 
corresponding wave functions. Operator product expansion therefore
corresponds to expanding the wave function 
$\Psi(j_2;x_2,\bx_2|h)\Psi(j_1;x_1,\bx_1|h)$ in terms of the basis given by
the $\Psi(j;x,\bx|h)$. The crucial observation in this context is that 
there exists a range of values for $j_1$, $j_2$ given by
\begin{equation}
|\Re(j_1+j_2+1)|<\fr{1}{2}\qquad\qquad |\Re(j_1-j_2)|<\fr{1}{2}
\end{equation}
where the product wave function 
$\Psi(j_2;x_2,\bx_2|h)\Psi(j_1;x_1,\bx_1|h)$ is normalizable.  
This range evidently includes the case where $\Phi^{j_2}$ and $\Phi^{j_1}$
correspond to the spectrum ($\Re(j_i)=-1/2$) and may be visualized as some
``strip'' around the axis $\Re(j_i)=-1/2$. By the completeness of the
basis spanned by the $\Psi(j;x,\bx|h)$ one may therefore expand 
\begin{equation}\label{exp} \Phi^{j_2}(x_2,\bx_2)\Phi^{j_1}(x_1,\bx_1)
=\int_{\CP_+} dj \int d^2x 
\;\,D\left(\begin{smallmatrix}j & j_2 & j_1 \\
x & x_2 & x_1 \end{smallmatrix} \right)\Phi^j(x,\bx).
\end{equation} 
Of course, taking the two point function with $\Phi^{-j-1}$,
$j\in\CP_+$ identifies 
the coefficients $D(\ldots)$ with the three point function: 
\[ D\left(\begin{smallmatrix}j & j_2 & j_1 \\
x & x_2 & x_1 \end{smallmatrix} \right)=
<\Phi^{-j-1}(x,\bx)\Phi_2(x_2,\bx_2)\Phi_1(x_1,\bx_1)>,\qquad j\in\CP_+ \]
In order to establish the relation to the bootstrap formalism
one should expand the coefficient $C$ for $x_2$ near $x_1$. The integration
over $x$ is then carried out by means of formula \rf{reflec}. 
One indeed recovers \rf{OPE1} and \rf{CCexp}.

One way to go beyond the considered region of values for the $j_i$ is by 
analytic continuation of \rf{exp}. This requires extending the integration 
in \rf{exp} to an integration over the whole axis $\CP\equiv -1/2+i\BR$, 
which may be done by using \rf{reflec}: 
\begin{equation}\label{exp2} \Phi^{j_2}(x_2,\bx_2)\Phi^{j_1}(x_1,\bx_1)
=\frac{1}{2i}\int_{\CP} dj \int d^2x 
\;\,D\left(\begin{smallmatrix}j & j_2 & j_1 \\
x & x_2 & x_1 \end{smallmatrix} \right)
\,\Psi^{j}(x,\bx),
\end{equation}
Analytically continuing \rf{exp2} in the 
parameter $j_1+j_2$ from $j_1+j_2<-1/2$ around the pole at $j_1+j_2=-1/2$
to $-1/2<j_1+j_2<0$ one encounters the situation that a pole of the integrand 
hits the integration contour. Suitably deforming it, one rewrites the integral 
as the sum of an integral over the original contour plus a 
residue contribution:
\begin{equation} \label{resterm}
\Phi^{j_2}(x_2,\bx_2)\Phi^{j_1}(x_1,\bx_1)
=\int d^2x E\left({}^{j_2}_{x_2}{}^{j_1}_{x_1};x\right)
\Phi^{-j_2-j_1-1}(x,\bx)
+\int_{\CP} dj\int d^2x \;\,D\left(\begin{smallmatrix}j & j_2 & j_1 \\
x & x_2 & x_1 \end{smallmatrix} \right)
\,\Phi^{j}(x,\bx),
\end{equation}
where
\[ E\left({}^{j_2}_{x_2}{}^{j_1}_{x_1};x\right)
=\mbox{Res}_{j=-j_1-j_2-1}{\left(\begin{smallmatrix}j & j_2 & j_1 \\
x & x_2 & x_1 \end{smallmatrix} \right)}
=(2j_2+1)(2j_1+1)|x-x_2|^{4j_2}|x-x_1|^{4j_1} \]

There are two further ways to understand this pattern of decomposition: 
First one may note that the product $\Psi_2\Psi_1$ is no longer normalizable 
for $-1/2<j_{21}<0$, due to the leading asymptotics given by 
$|v|^{2(j_1+j_2)}$. However, the first term on the 
right hand side of \rf{resterm} can be
seen\footnote{Again most easily by expanding in $x_2-x_1$ and using 
\rf{reflec}} to have the same leading 
asymptotics, substracting it from the  
left hand side will then yield something normalizable that 
can be expanded in terms of the $\Phi^j$ with $j\in \CP$.
One learns that even certain non-normalizable states possess a well-defined
expansion if the basis $\{\Phi^j; j\in \CP_+\}$ is suitably extended. 

Second, one may note that if $j_i\in\BR$, $-1<j_i<0$ then  
the operators $\Psi_i$ transform as unitary representations of the 
supplementary series under $SL(2,\BC)$. The found pattern of decomposition is
precisely that predicted by Naimark's theorem on the decomposition of tensor
products of $SL(2,\BC)$ representations \cite{N}.

A few remarks are in order:
\begin{enumerate}
\item Needless to say that the
decomposition for general $j_1$, $j_2$ can be found by further  
analytic continuation, picking up more and more residue terms. 
In some cases the OPE reduces to a sum over finitely many $j$: 
This will happen iff either $2j_1+1\in\BZ^{>0}$ or $j_2+1\in\BZ^{>0}$ ,
which is the 
case where one of the $\Psi^{j_i}$ $i=1,2$ corresponds to a finite dimensional
representation of $SL(2,\BC)$.
\item Nonvanishing of the three-point function does not imply appearance of 
any one of the three fields in the operator product expansion of the 
other two.
\end{enumerate}

\subsection{Four point function; factorization}

Start by considering the four point function $<\Psi_4\ldots\Psi_1>$ in 
the case that 
\begin{equation}
\begin{aligned}
|\Re(j_1+j_2+1)|<&\fr{1}{2}\\ 
|\Re(j_3+j_4+1)|<&\fr{1}{2}
\end{aligned}\qquad
\begin{aligned}
|\Re(j_1-j_2)|<& \fr{1}{2}\\
|\Re(j_3-j_4)|<& \fr{1}{2}
\end{aligned}\label{fundreg}\end{equation}
Under these conditions the product $\Psi_2\Psi_1$ 
is square-integrable, and can be expanded according to \rf{exp}. 
This yields the following representation for the four point function: 
\[ <\Phi_4\ldots \Phi_1>=-i\int_{\CP_+} dj \int d^2x <\Phi_4\Phi_3
\Phi_{j}(x,\bx)><\Phi^{-j-1}(x,\bx)
\Phi_2\Phi_1>.\] 
The integral over $x$ can be performed with the result
\begin{eqnarray*} \lefteqn{\int d^2x <\Phi_4\Phi_3\Phi_{j}(x,\bx)>
<\Phi^{-j-1}(x,\bx)\Phi_2\Phi_1>= }\\
& = & |x_{43}|^{2(j_4+j_3-j_2-j_1)}|x_{42}|^{4j_2}|x_{41}|^{2(j_4+j_1-j_3-j_2)}
|x_{31}|^{2(j_3+j_2+j_1-j_4)} \\[.5ex]
& & \Big( D_j \; |x|^{2(j_1+j_2+j+1)}\; F_{-j-1}(x)F_{-j-1}(\bx)
+ E_j \; |x|^{2(j_1+j_2-j)}\; F_{j}(x)F_{j}(\bx)\Big), 
\end{eqnarray*}
where $x_{ij}=x_i-x_j$, $x=\frac{(x_2-x_1)(x_4-x_3)}{(x_3-x_1)(x_4-x_2)}$,
\begin{equation}\begin{split}
D_j = & C(j_4,j_3,j)C(-j-1,j_2,j_1)
\frac{\ga(j+1+j_4-j_3)}{\ga(j_4-j_3-j)\ga(2j+2)} \\
 = & \frac{1}{2j+1}C(j_4,j_3,-j-1)C(-j-1,j_2,j_1), \displaybreak[0] \\
E_j = & C(j_4,j_3,j)C(-j-1,j_2,j_1)
\frac{\ga(-j+j_2-j_1)}{\ga(j+1+j_2-j_1)\ga(-2j)} \\
 = & -\frac{1}{2j+1}C(j_4,j_3,j)C(j,j_2,j_1)=D_{-j-1}\displaybreak[0] \\
F_j(z)  = & F(j_4-j_3-j,j_1-j_2-j;-2j;z)\end{split}\end{equation}
One thereby finds an expansion into conformal blocks of the form 
\rf{s-channel}.  

Of course one can treat the case of more general values for the $j_i$ by 
analytic continuation, which will again lead to a sum over residue terms.
However, if one considers analytic continuation to a region where 
the conditions on $j_3,j_4$ in \rf{fundreg} are violated, one will get 
residue terms that correspond to non-normalizable operators in the 
product $\Phi_4\Phi_3$. Proper description of such contributions 
will be somewhat problematic in a canonical operator formalism.  
\newpage

\section{Appendix A: Spectral decomposition}

A (plane wave) basis for $L^2(H_3^+,dh)$ may be found by diagonalizing 
a complete set of commuting differential operators. A convenient choice is
to take the operators 
\begin{equation}\begin{array}{c}\dst
K^0\equiv \frac{1}{2}(J^0-\bJ^0)=-v\frac{\pa}{\pa v}+\bv\frac{\pa}{\pa \bv}, 
\qquad F^0\equiv \frac{1}{2}(J^0+\bJ^0)=-i\frac{\pa}{\pa \phi} \\[2ex]
\begin{array}{rcl}Q&=&\dst
\frac{1}{2}( 2(  J^0)^2+  J^+  J^-+  J^-  J^+)=
\frac{1}{2}( 2(\bJ^0)^2+\bJ^+\bJ^-+\bJ^-\bJ^+) \\[2ex]
 &=& \dst (1+|v|^2)\frac{\pa^2}{\pa v\pa\bv}+\frac{1}{4} 
\left( v\frac{\pa}{\pa v}-\bv\frac{\pa}{\pa \bv}\right)^2+
\frac{1}{2}\left(v\frac{\pa}{\pa v}+\bv\frac{\pa}{\pa \bv}\right)
+\frac{1}{4}\frac{1}{1+|v|^2}\frac{\pa^2}{\pa \phi^2}
\end{array}\end{array}\end{equation}
By writing $v$ as $v=e^{i\vf}\sqrt{y}$, $\vf\in[-\pi,\pi]$, $y\in\BR$ one has 
$K^0=-i\frac{\pa}{\pa \vf}$, so $K^0$ has spectrum $\BZ$, whereas $F^0$
has spectrum $\BR$. On an eigenspace of $K^0$, $F^0$ with eigenvalues 
$n$, $p$ the operator $Q$ reduces to
\begin{equation} 
Q_{np}=\frac{\pa}{\pa y} y(1+y)\frac{\pa}{\pa y}
-\frac{n^2}{y}-\frac{p^2}{1+y} 
\end{equation} 

\subsection{Self-adjointness of $Q_{np}$}

One has
\begin{equation} \int_{0}^{\infty}dy (Q_{np}f(y))^*g(y)-
\int_{0}^{\infty}dy (f(y))^*Q_{np}g(y)=
\left[y(1+y)
\left(f\frac{\pa}{\pa y}g-g\frac{\pa}{\pa y}f\right)\right]_{0}^{\infty},
\label{symm?}\end{equation}
so that $Q_{np}$ is symmetric on the dense subspace $D$ of $L^2(\BR^{\geq 0})$
that consists of functions regular at $0$ and $\infty$. According to the 
general theory of self-adjoint extensions of unbounded symmetric operators
\cite{AG} one needs to know whether there exist normalizable eigenfunctions
to eigenvalues with strictly positive or strictly negative imaginary part.
The eigenvalue equation $Q_{np}f=j(j+1)f$ is transformed into the 
hypergeometric differential equation by means of $f=y^n(1+y)^{ip}g$.
The solutions with specified asymptotic behavior near $y=\infty$ are
$g(y;j)$ and $g(y,-j-1)$, where
\[ g(y;j)=y^{j}(1+y^{-1})^{ip}
F(\fr{1}{2}(|n|+ip)-j,\fr{1}{2}(-|n|+ip)-j;-2j;-y).  \]
Necessary for $g(y;j)$ to be square-integrable is $\Re(j)<-\frac{1}{2}$.
However, $g(y;j)$ behaves for $y\ra 0$ as $a_n y^{n}+b_n y^{-n}$ 
with $a_n,b_n\neq 0$ for $n\neq 0$, cf. \cite{E}, p.109, eqn. (7). It is 
therefore not square-integrable for $n\neq 0$. It remains to consider $n=0$.
In that case one should observe that $g(y;j)\sim \log(y)$ for $y\ra 0$, 
which leads to a nonzero contribution on the r.h.s.of (\ref{symm?}) if 
$f\in D$. It is therefore not possible to include $g(y;j)$ in an extension 
$D'$ of the domain $D$ such that $Q_{0p}$ becomes selfadjoint on $D'$. 
By the general theory of \cite{AG} one therefore has a unique selfadjoint 
extension of $(Q_{np},D)$. 

\subsection{The resolvent}

In order to determine the spectrum of $Q_{np}$ it is useful to construct
its resolvent $R(q)=(Q_{np}-q)^{-1}$; $q\in\BC$: 
The spectrum is essentially encoded in the analyticity properties of $R(q)$
\cite{Yo}. Poles on the real axis correspond to eigenvalues, cuts to
the continuous spectrum.

For any differential operator of the form $\CD_y\equiv \pa_yp(y)\pa_y-r(y)$ 
one may construct the kernel of the resolvent $\CR\equiv
(\CD-q)^{-1}$ in the form
\[ R(y,y';q)=N^{-1}\Big(\Theta(y-y')g(y;q)f(y';q)+
\Theta(y'-y)f(y;q)g(y';q)\Big), \]
where $f(y;q)$ and $g(y,q)$ are two linearly independent solutions of 
$(\CD_y-q)F=0$ and $\Theta(x)=0$ for $x<0$, $\Theta(x)=1$ for $x>0$.
Indeed, straightforward calculation shows that 
\[ (\CD_y-q)R(y,y';q)=N^{-1}p(f\pa_yg-g\pa_yf)\de(y-y').\]
Moreover, the combination $p(f\pa_yg-g\pa_yf)$ is constant as a consequence 
of $(\CD_y-q)F=0$, $F=f,g$, so that the choice $N=p(f\pa_yg-g\pa_yf)$ indeed 
gives the resolvent kernel.

In the present case it is useful to parameterize the eigenfunctions 
in terms of the variable $j$, which from now on will have to be considered 
as a function of the eigenvalue $q$ defined by 
\begin{equation}\label{root}
j\equiv -\frac{1}{2}+\sqrt{\frac{1}{4}+q}\mbox{ for }
q\in\BC\setminus (\infty,-1/4],\quad
j\equiv -\frac{1}{2}+i\sqrt{-\frac{1}{4}-q}\mbox{ for }
q\in (\infty,-1/4].
\end{equation}
$f(y;q)$ and $g(y,q)$ will now be chosen as
\begin{eqnarray*}
f(y;j)&=&y^{|n|}(1+y)^{ip}\;
F\Big(\fr{1}{2}(|n|+ip)-j,\fr{1}{2}(|n|+ip)+j+1;1+|n|;-y\Big) \\
g(y;j)&=&B_{np}(j) \; y^{j}(1+y^{-1})^{ip}\;
F\Big(\fr{1}{2}(|n|+ip)-j,\fr{1}{2}(-|n|+ip)-j;-2j;-\fr{1}{y}\Big)\\
& & B_{np}(j)\equiv \frac{\Ga(1+|n|)\Ga(2j+1)}{\Ga(\frac{1}{2}(|n|+ip)+j+1)
\Ga( \frac{1}{2}(|n|-ip)+j+1)}. 
\end{eqnarray*}
The factor $B_{np}(j)$ has been chosen for convenience since now the 
standard connection formula for the hypergeometric functions (\cite{E}, p. 108,
eqn. (1)) reads $f(y,j)=g(y,j)+g(y,-j-1)$. The normalization is therby 
evaluated as
\begin{eqnarray*} 
N=\lim_{y\ra\infty} p(f\pa_yg-g\pa_yf)&=&\frac{1}{2j+1}B_{np}(j)B_{np}(-j-1)
\end{eqnarray*}

\subsection{The spectrum}

Having explicitely constructed the resolvent it is easy to 
determine the spectrum: According to \cite{Yo}, chap. XI, sect.9 
the discrete part of the
spectrum would show up as poles of the resolvent $R(q)$ on the real $q-$axis.
There are none in the present case. Furthermore, the 
continuous spectrum corresponds to jumps of $\CR(q)$ on the 
real axis as is manifest in the formula for the spectral projection
onto the interval $[a,b]$ given in
(\cite{Yo}, loc. cit.): 
\begin{equation}\label{specproj}\CP_{[a,b]}v =
\lim_{\ep\ra 0+}\frac{1}{2\pi i}\left[
\int_{a}^{b}dq\;\CR(q+i\ep)v-\int_{a}^{b}dq \;\CR(q-i\ep)v
\right]. \end{equation}
If and only if the resolvent $\CR$ has a jump at a certain value 
$q\in\BR$ one will get a contribution to (\ref{specproj}), so $q$ belongs to
the continuous spectrum. Here the jump arises from the square-root branch cut
in (\ref{root}): One has $\lim_{\ep\ra 0+}j(q+i\ep)=j(q)$ and 
$\lim_{\ep\ra 0+}j(q-i\ep)=-j(q)-1$ for $q\in(-\infty,-1/4]$ and no jump 
otherwise.
The continuous spectrum is therefore found to be $(-\infty,-1/4]$. 

This may be reformulated as a completeness relation by noting that 
$\CP_{(-\infty,-1/4]}=Id$, so rewriting (\ref{specproj}) in terms of the
corresponding kernels gives
\[ \begin{array}{rcl}
\de(y-y') 
= & \dst \frac{1}{2\pi i}\int_{-\infty}^{-\frac{1}{4}}
\frac{dq}{2j+1}\frac{1}{|B_{np}(j)|^2}\bigg\{ & \Theta(y-y')
\Big( g(y;j)+g(y,-j-1)\Big)f(y';j) \\[2ex]
& & \Theta(y'-y)
\Big( g(y';j)+g(y',-j-1)\Big)f(y;j)\bigg\}\\
= & \multicolumn{2}{l}{ 
\dst \frac{1}{2\pi i}\int_{\CP_+}dj \;\;\frac{1}{|B_{np}(j)|^2}\;\; 
f(y,j)f(y',j) \qquad \qquad\CP_+\equiv -\fr{1}{2}+i\BR^{\geq 0}}
\end{array} \]

\subsection{Basis I}

The results of the previous subsections show that the set of functions
\begin{equation}\label{BI}
\Psi^{j}_{np}(h)=B_{np}^{-1}(-j-1)e^{in\vf}e^{ip\phi}f(y,j)
\qquad n\in\BZ,p\in\BR,j\in \CP_+ \end{equation}
is complete in $L^2(H_3^+)$.
One may also check that it is $\de$-function orthonormalized:
The scalar product may be evaluated by means of the formula
\begin{eqnarray*} \lefteqn{
(j(j+1)-j'(j'+1))\int_{0}^{\infty}dy \;\; f(y;j)
f(y;j') }\\
& = & \lim_{y\ra\infty} y(1+y)\left(
f(y;j')\pa_y f(y;j)-f(y;j)\pa_y f(y;j')\right),\end{eqnarray*}
which follows from the fact that $f(y;j)$ and 
$f(y;j')$ are eigenfunctions of the differential operator $Q_{np}$
with eigenvalues $j(j+1)$ and $j'(j'+1)$ respectively.
The result is 
\begin{equation} <\Psi^j_{np},\Psi^{j'}_{n'p'}>=
\int_{H_3^+}dh \left(\Psi^j_{np}(h)\right)^{\ast}\Psi^{j'}_{n'p'}(h)=
(2\pi)^3\delta_{n,n'}\delta(p-p')i\delta(j-j'), \label{ortho}\end{equation}
where $j,j'$ are assumed to be in $\CP_+$.

Finally note that alternatively one might take the $\Psi^j$ with 
$j\in -1/2+i\BR^{\leq 0}$ as basis as they differ only by a phase 
factor
\begin{equation}\label{reflI}
\Psi^j_{np}(h)=\frac{B_{np}(j)}{B_{np}(-j-1)}\Psi^{-j-1}_{np}(h).
\end{equation}

\subsection{Basis II}

A second useful basis may be introduced as the following Fourier transform 
of the basis \rf{BI}
\[ \Psi(j,x,\bx|h)=\frac{1}{(2\pi)^2}\sum_{n\in\BZ}\int_{\BR}dp 
\;\; e^{-in\arg(x)}|x|^{2j-ip}\Psi^{j}_{np}(h). \]
In order to see that the result is simply 
\[ \Phi^j(x,\bx|h)=\frac{2j+1}{\pi}
\left( (x,1)\cdot h\cdot \binom{\bx}{1}\right)^{2j} 
\] 
one may i.e. use the following reasoning: First note that  
$\Psi^j(x,\bx|h)$ satisfies an intertwining property with respect 
to $SL(2,\BC)$-transformations:
\begin{equation}\label{intertw} \Psi^j(x,\bx|ghg^{\dagger})=
|\beta x+\delta|^{4j} \Psi\left(\frac{\al x+\ga}{\be x+\de},c.c.|h \right)
\quad\mbox{if}\quad
g=\left(\begin{array}{cc} \al & \be \\ \ga & \de \end{array} \right). 
\end{equation} 
On the infinitesimal level this amounts to a couple of differential equations 
relating derivatives w.r.t. $h$ to those w.r.t. $x$, in particular
($x=e^{i\xi}r$)
\[ \left(\frac{\pa}{\pa \vf}+\frac{\pa}{\pa\xi}\right)\Psi^j=0
\qquad 
\left(\frac{\pa}{\pa \phi}+r\frac{\pa}{\pa r}-2j\right)\Psi^j=0
\qquad Q\Psi^j=j(j+1)\Phi^j \]
By the inverse transform 
\begin{equation}\label{FI} \tilde{\Psi}^j_{np}(h)\equiv
\int_{\BC}d^2x \;\; e^{in\arg(x)}|x|^{-2j-2+ip} \Psi^j(x,\bx|h). \end{equation}
these differential equations turn into the eigenvalue equations
for $Q$, $F^0$, $K^0$. Since moreover $\Psi(j;x,\bx|h)$ is regular for 
$y\ra 0$ one finds that $\tilde{\Psi}^j_{np}(h)$ must be proportional to 
$\Psi^j_{np}(h)$. To find the constant of proportionality one may consider
the integral \rf{FI} in the limit $y\ra\infty$ where it can be reduced to
\begin{equation}\label{dotint}
\frac{2j+1}{\pi}\int d^2x e^{in\arg(x)}|x|^{-2j-2+ip}|x-1|^{4j}=
\frac{B_{np}(j)}{B_{np}(-j-1)},
\end{equation}
so that indeed $\tilde{\Psi}^j_{np}(h)=\Psi^j_{np}(h)$.

The relations expressing orthogonality and completeness of the 
$\Psi^j(x,\bx|h)$ are now easily obtained from those of the $\Psi^j_{np}(h)$:
\begin{eqnarray}
\int_{H_3^+} dh\;\; (\Psi(j;x,\bx|h))^*
\Psi(j';x',\bx'|h)&=& 2\pi\de^{(2)}(x-x')i\de(j-j') \\
\frac{1}{i}\int_{\CP_+}dj \int_{\BC}d^2x 
(\Psi(j;x,\bx|h))^*\Psi(j;x,\bx|h')&=&(2\pi)^3 \de(\vf-\vf')\de(\phi-\phi')
\de(y-y') \end{eqnarray}

The relation between $\Psi(j;x,\bx|h)$ and $\Psi(-j-1;x,\bx|h)$ 
is obtained by Fourier transform of \rf{reflI} and again using \rf{dotint}: 
\begin{equation} \Psi^j(x,\bx)=\frac{2j+1}{\pi}\int\! d^2x'\; |x-x'|^{4j}
\Psi^{-j-1}(x',\bx').\label{reflII} \end{equation} 

\section{Appendix B: Fusion relations}

I will now present a direct calculation inspired by \cite{BM} of the 
semi-classical fusion matrix.
This approach will have the additional advantage to work for a larger 
range of the $j_i$-values, so I will consider complex $j_i$ with certain
restrictions to be specified below only on their real parts.

The basic ingredient is the following formula of Burchnall and Chaundy: 
\[
(1-x)^{d}=
\sum_{n=0}^{\infty}  (-)^n\frac{(a)_n(b)_n}{(c+n-1)_n n!}\;
{}_3 F_2\binom{-d,c+n-1,-n}{a \quad b} 
\; x^nF(a+n,b+n;c+2n;x) 
\]
This formula holds for any value of $a,b,c$ on the right hand side. It can be 
turned into an expansion into eigenfunctions of $P$ if one identifies
\[ \begin{array}{ccc}
a+n=j_1-j_2-j & \qquad & b+n=j_4-j_3-j \\[.5ex]
c+2n=-2j & \qquad & n=j_1+j_2-j.  
\end{array}\]
Summation over $n$ may be traded for integration over $j$'s corresponding 
to principal series intermediate representations by identifying the 
right hand side as the sum over the residues at $j-j_1-j_2=-n$ of 
\begin{eqnarray*}
H(j;j_4,\ldots,j_1;d;x) &\equiv &
\Ga(j-j_1-j_2)\frac{\Ga(j_1-j_2-j)\Ga(j_4-j_3-j)\Ga(-j_1-j_2-j-1)}
{\Ga(-2j_2)\Ga(j_4-j_3-j_2-j_1)\Ga(-2j-1)}\ti \\
& & 
\ti{}_3 F_2\binom{-d,-j-j_1-j_2-1,j-j_1-j_2}{ 
-2j_2\quad j_4-j_3-j_2-j_1} \\
& & \ti x^{j_1+j_2-j}F(j_1-j_2-j,j_4-j_3-j;-2j;x).
\end{eqnarray*}
In fact, considered as function of $j$ one finds that $H(j_4,\ldots,j_1;j;x)$
has the following poles:
\[ \begin{array}{ccc}
j=j_1+j_2-n_1 & \quad & j=-j_2-j_2-1+n_2 \\[.5ex]
j=j_1-j_2+n_3 & \quad & j=j_4-j_3+n_4 
\end{array} \quad n_1,\ldots,n_4\in\BZ^{\geq 0} 
\]
One finds only the ``wanted'' poles $j=j_1+j_2-n; n=0,1,2,\ldots$ within the 
half-plane $\{ j; \Re(j)<-\frac{1}{2}\}$ iff
\[ 
\Re(j_1+j_2)<-\frac{1}{2}  \qquad  \Re(j_1-j_2)>-\frac{1}{2} \qquad 
\Re(j_4-j_3)>-\frac{1}{2} 
\]
In this case one may rewrite the sum over residues as limit of the 
integrations over the closed contour
\[ \CC_r\equiv \left\{ j=-\fr{1}{2}+i\si;\si\in\BR \right\}\cup
\left\{ j ; \left|j+\fr{1}{2}\right|=r, \Re(j+\fr{1}{2})<0 \right\} \]
for $r\ra\infty$. However, by using the estimates on the integrand given in
\cite{BM}, one recognizes that the contributions from the semi-circle
vanish for $r\ra\infty$. One is left with 
\begin{equation}\label{SW}
(1-x)^{d}  \dst =\int_{-\frac{1}{2}-i\infty}^{-\frac{1}{2}+i\infty}dj
\; H(j;j_4,\ldots,j_1;d;x).
\end{equation}

Now consider the semi-classical t-channel conformal block 
\[ F^t_{j_{32}}=(1-x)^{j_2+j_3-j_{32}}
F(j_3-j_2-j_{32},j_4-j_1-j_{32};-2j_{32};1-x) \]
By expanding the hypergeometric function as power series in $1-x$, applying
(\ref{SW}) and exchanging summation with integration one arrives at
\[ F^t_{j_{32}}(x)=\int_{-\frac{1}{2}-i\infty}^{-\frac{1}{2}+i\infty}dj_{21}
F_{j_{32}j_{21}}\!\left[{}^{j_3j_2}_{j_4j_1}\right] 
F^s_{j_{21}}(x),\]
where
\[
\begin{array}{rccl}\dst
F_{j_{32}j_{21}}\!\left[{}^{j_3j_2}_{j_4j_1}\right] &=&
\multicolumn{2}{l}{\dst
\frac{\Ga(j_{21}-j_1-j_2)
\Ga(j_4-j_3-j_{21})\Ga(-j_{21}-j_1-j_2-1)\Ga(j_1-j_2-j_{21})}{\Ga(-2j_{21}-1)
\Ga(-2j_2)\Ga(j_4-j_3-j_2-j_1)}} \\[2ex]
& & \ti \dst
\sum_{n=0}^{\infty} & \dst
\frac{\Ga(j_3-j_2-j_{32}+n)\Ga(j_4-j_1-j_{32}+n)\Ga(-2j_{32})}
{\Ga(j_3-j_2-j_{32})\Ga(j_4-j_1-j_{32})\Ga(-2j_{32}+n) \;n!}\\
& & & \dst
\ti {}_3 F_2\binom{
j_{32}-j_3-j_2-n,-j-j_1-j_2-1,j-j_1-j_2}{ 
-2j_2\quad j_4-j_3-j_2-j_1}
\end{array}
\]
The summands have large $n$ asymptotics $n^{j_{21}+j_3+j_4}$, 
so if also $\Re(j_3+j_4)<-1/2$
one has absolute convergence, which justifies the exchange of   
summation with integration that had been performed.

\newcommand{\CMP}[3]{{\it Comm. Math. Phys. }{\bf #1} (#2) #3}
\newcommand{\LMP}[3]{{\it Lett. Math. Phys. }{\bf #1} (#2) #3}
\newcommand{\IMP}[3]{{\it Int. J. Mod. Phys. }{\bf A#1} (#2) #3}
\newcommand{\NP}[3]{{\it Nucl. Phys. }{\bf B#1} (#2) #3}
\newcommand{\PL}[3]{{\it Phys. Lett. }{\bf B#1} (#2) #3}
\newcommand{\MPL}[3]{{\it Mod. Phys. Lett. }{\bf A#1} (#2) #3}
\newcommand{\PRL}[3]{{\it Phys. Rev. Lett. }{\bf #1} (#2) #3}
\newcommand{\AP}[3]{{\it Ann. Phys. (N.Y.) }{\bf #1} (#2) #3}
\newcommand{\LMJ}[3]{{\it Leningrad Math. J. }{\bf #1} (#2) #3}
\newcommand{\FAA}[3]{{\it Funct. Anal. Appl. }{\bf #1} (#2) #3}
\newcommand{\PTPS}[3]{{\it Progr. Theor. Phys. Suppl. }{\bf #1} (#2) #3}
\newcommand{\LMN}[3]{{\it Lecture Notes in Mathematics }{\bf #1} (#2) #2}

\end{document}